\documentclass[slac_one]{revtex4}
\usepackage{graphicx}
\usepackage{fancyhdr}
\usepackage{amsmath,epsf}
\def\Vud  {$\mathrm{|V_{ud}|}$}
\def\Vus  {$\mathrm{|V_{us}|}$}
\def\Vub  {$\mathrm{|V_{ub}|}$}
\def\Vcd  {$\mathrm{|V_{cd}|}$}
\def\Vcs  {$\mathrm{|V_{cs}|}$}
\def\Vcb  {$\mathrm{|V_{cb}|}$}
\def\Vtd  {$\mathrm{|V_{td}|}$}
\def\Vts  {$\mathrm{|V_{ts}|}$}
\def\Vtb  {$\mathrm{|V_{tb}|}$}
\def\to   {\rightarrow}
\def\belle {Belle}
\newcommand{\tev}{\ensuremath{\mathrm{\,Te\kern -0.1em V}}\xspace}
\newcommand{\gev}{\ensuremath{\mathrm{\,Ge\kern -0.1em V}}\xspace}
\newcommand{\mev}{\ensuremath{\mathrm{\,Me\kern -0.1em V}}\xspace}
\newcommand{\kev}{\ensuremath{\mathrm{\,ke\kern -0.1em V}}\xspace}
\newcommand{\ev}{\ensuremath{\mathrm{\,e\kern -0.1em V}}\xspace}
\usepackage{relsize}
\newcommand{\eg}{{\em e.g.}}
\newcommand{\el}{$E_\ell$}
\newcommand{\mx}{$M_X$}

\newcommand{\smax}{$s^{max}$}
\newcommand{\pplus}{$P^+$}
\def\babar{\mbox{\slshape B\kern-0.1em{\smaller A}\kern-0.1em
    B\kern-0.1em{\smaller A\kern-0.2em R}}}

\begin{document}

\title{A Review of the Magnitudes of the CKM Matrix Elements} 

%

\author{Francesca Di Lodovico}
\affiliation{Queen Mary, University of London, E1 4NS, London, UK}

\begin{abstract}
Flavour mixing is described within the Standard Model by
the Cabibbo--Kobayashi--Maskawa matrix elements.
With the increasingly higher statistics collected by many experiments,
the matrix elements are measured with improved precision, allowing
for more stringent tests of the Standard Model.
In this paper, a review of the current status of the
absolute values of the CKM matrix elements is presented, with particular 
attention to the latest measurements.
\end{abstract}

\maketitle

\thispagestyle{fancy}


\section{INTRODUCTION} 
The Standard Model (SM) accounts for flavour--changing quark 
transitions in terms of a V--A charged weak current operator $\cal J^\mu$ that couples to the
$W$ boson according to the interaction Lagrangian:
${\cal L}_{int}= - \dfrac{g}{\sqrt{2}}(\cal{J}^\mu W^+_\mu + \cal{J}^{\mu +} W^-_\mu )$,
where for quark transitions
${\cal J}^\mu = \Sigma_{i,j} V_{i,j}J^\mu_{ij} = \Sigma_{i,j}\overline{u}_i\gamma^\mu\dfrac{1}{2}(1-\gamma_5)V_{ij}d_j$ ($i,j$ run over the three quark generations).
The $V_{i,j}$ are the CKM matrix elements. The field operator 
$u_i$ ($d_j$) annihilates the $u, c, t$ ($d, s, b$)
quarks. The operator $\cal W^+_\mu$
annihilates a $W^+$ or creates a $W^-$. The reverse is true for $\cal W^-_\mu$. Thus, the CKM
matrix $V$ can be regarded as a rotation of the quark mass eigenstates $d, s, b$
to a new set $d^\prime, s^\prime, b^\prime$ with diagonal coupling to $u, c, t$.
The standard notation to represent it is:
\begin{equation}
\begin{pmatrix}
d' \\ s' \\ b' 
\end{pmatrix} 
=
\begin{pmatrix}
V_{ud} &V_{us} &V_{ub} \\
V_{cd} &V_{cs} &V_{cb} \\ 
V_{td} &V_{ts} &V_{tb} \\
\end{pmatrix}
\begin{pmatrix}
d \\ s \\ b \\
\end{pmatrix}
\end{equation}
which is an {\it almost} unitarity matrix. However, none of the off-diagonal elements are
exactly zero, so generation changing transitions between quarks are possible.
The values of the CKM matrix elements are fundamental parameters of the SM and cannot be predicted.
In the following, a review of the current values of the absolute values of the 
CKM matrix elements is presented. For each matrix element, all 
measurements which lead to the measurement of that matrix element are presented.
A detailed description of the measurements is only given for the
most recent results.

\section{FIRST ROW OF THE CKM MATRIX}
\subsection{\boldmath\Vud}
Three main techniques lead to the measurement of \Vud, with differing precision. 
The most precise determination comes from 
superallowed $0^+\to 0^+$ nuclear $\beta$ transitions.
The intensity of any $\beta$ transition is expressed as an $ft$ value, which depends
on the transition energy ($Q_{EC}$), the half life of the $\beta$ emitter, and the 
branching ratio for the particular transition of interest. 
Thirteen superallowed transition have an $ft$ value measured to a precision
between $0.03\%$ and $0.3\%$~\cite{Towner:2007np}.
From the experimentally determined $ft$ value,
applying calculated transition correction terms in the SM, the corrected quantity
${\cal F}t$~\cite{Towner:2007np}, should be identical for all cases. From ${\cal F}t$, 
\Vud\ can be extracted.  
An improved measurement of the $Q_{EC}$ values of $\rm ^{50}$Mn 
and $\rm ^{54}$Co~\cite{Eronen:2007qc}, which
removes the discrepancy between the $\cal{F}$$t$  values of those two elements and the 
averages $\cal{F}$$t$ value of the most precise thirteen superallowed 
transitions~\cite{Towner:2007np}, leads to a value of \Vud = $0.97408(26)$.
It has a precisions of $0.03\%$ and its error is dominated by the theoretical
uncertainties on the radiative and isospin--symmetry--breaking corrections.

Moreover, from the knowledge of the neutron lifetime, $\tau_n$, and the axial--vector/vector couplings,
$g_A\equiv G_A/G_V$, another measurement of \Vud\ can be performed, leading to
\Vud = $0.9746(4)_{\tau_n}(18)_{g_A}(2)_{RC}$~\cite{Amsler:2008pdg},
where the errors are due to $\tau_n$, $g_A$ and $RC$, where $RC$ denotes the entire effects
of electroweak radiative corrections, nuclear structure, and isospin violating nuclear effects.
The error is dominated by the uncertainty on $g_A$.
Another measurement of $\tau_n$\cite{Serebrov:2004zf} leads to a higher value of \Vud.
Future studies are expected to resolve this inconsistency.

Finally, the pion decay $\pi^+\to\pi^0e^+\nu$ can be 
used to extract \Vud, as its rate depends on \Vud$^2$.
This approach is theoretically very clean, as it is free from nuclear structure
uncertainties, but the decay is disadvantaged by a low branching ratio. 
The most precise measurement is 
from PIBETA~\cite{Pocanic:2003pf} and gives \Vud = $0.9728(30)$, with an error of 0.3$\%$,
and in agreement with the above values.

\subsection{\boldmath\Vus}

The most precise way to measure \Vus\ comes from semileptonic kaon decays,
whose rate is proportional to \Vus$^2f_+(0)^2$, where $f_+(q^2)$ at $q^2=0$
is the $K^0\rightarrow \pi^+$ transition at zero moment transfer [$q^2=(p_K-p_\pi)^2=0$],
in the limit $m_u=m_d$ and $\alpha_{em}\rightarrow 0$.
Using worldwide averages of lifetimes, branching ratios, phace space integrals,
and the radiative and $SU(2)$ breaking corrections
for $K_L\rightarrow \pi e\nu$, $K_L\rightarrow \pi \mu \nu$, 
$K_s\rightarrow \pi e\nu$, $K^\pm\rightarrow \pi e\nu$ and $K^\pm\rightarrow \pi \mu\nu$,
globally indicated as $K_{\ell 3}$, from KTeV, NA48, KLOE and ISTRA+, 
the FlaviaNet Kaon group~\cite{Antonelli:2008jg} 
extracts \Vus$f_+(0) = 0.21664(48)$, which has a precision of $\sim 0.2\%$.
Differences between the current measurements of the branching fractions 
and the past ones depend on a proper treatment of the radiative effects.
Choosing $f_+(0)=0.964(5)$ from UKQCD-RBC~\cite{Antonio:2007mh}, the value of \Vus\ is 
\Vus $= 0.2246\pm 0.0012$, where the dominant error is experimental.
Note that indirect tests of the form factor $f_+(0)$ from the $K_{3\ell}$
decays requires a further understanding of the disagreement between the NA48 result~\cite{Lai:2007dx}
and the other measurements.

Moreover, using the ratio of kaon and pion leptonic decays 
$K^+\rightarrow \mu^+\nu/\pi\rightarrow \mu^+\nu$, whose kaon rate
is dominated by KLOE, FlaviaNet quotes 
\Vus$/$\Vud$f_K/f_\pi = 0.2760\pm0.0006$~\cite{Antonelli:2008jg}, where
$f_K/f_\pi$ is the ratio of the kaon and pion decay constants 
in the limit $m_u=m_d$ and $\alpha_{em}\rightarrow 0$.
Using the form factor ratio from the MILC-HPQCD collaboration 
$f_K/f_\pi = 1.189(7)$~\cite{Follana:2007uv}, the value 
\Vus$/$\Vud $ = 0.2321\pm 0.0015$ is obtained, where the accuracy is 
limited by the knowledge of the decay constants.\\
These results together with the measured value of \Vud\ from superallowed $\beta$ decays, 
can be used to test the unitarity of the first row of the CKM matrix~\cite{Antonelli:2008jg},
leading to 0.9999(9), neglecting \Vub\ with $\chi^2/ndof$=0.65/1.

Measurements of hadronic tau decays provide a new test of the value
of the CKM matrix element \Vus, although affected by a higher statistical error
with respect to the kaon decays. 
The extraction of \Vus\ from hadronic tau decays involves moments of the
invariant mass distributions of the final states hadrons~\cite{Le Diberder:1992fr}.
The averaged value of
\Vus\ is \Vus $=0.2159 \pm 0.0030$~\cite{Banerjee:2008hg}, 
where the dominant contribution error is due to the
experimental uncertainty. The average does not include
correlations between different measurement as they are not available yet. 
A recent discussion on the size
of the theoretical errors can be found in Ref.~\cite{Maltman:2008ib}.
The \Vus\ value from tau decays is about 3$\sigma$ lower than the value extracted from
kaon decays. 
A new result presented by \babar\ on the ratio of the
branching fractions for $\tau^-\rightarrow K^-\nu$ and $\tau^-\rightarrow \pi^-\nu$ 
decays gives \Vus $= 0.2256 \pm 0.0023$~\cite{Banerjee:2008hg}, 
using $f_K/f_\pi$ from Ref.~\cite{Follana:2007uv}
and \Vud\ from Ref.~\cite{Eronen:2007qc}, whose error is dominated by uncertainties
in particle identification. 
Although, this value of \Vus\ is consistent with the one from kaon decays,
individually both branching fractions are lower than the universality predictions.
More understanding both of the experimental (e.g. correlations in the
averages) and theoretical issues is needed to solve the current discrepancy.

Finally, hyperon decays can lead to a further determination of \Vus.
The semileptonic decay of a spin $1/2$ hyperon involves the hadronic matrix
elements of the vector and axial--vector currents. Supported by the fact that there
are no first--order corrections to the vector form factor~\cite{Ademollo:1964sr}, and using
an experimental measurement of the axial and vector form factor ratio, thus
avoiding SU(3) breaking effects, a value of 
\Vus $ = 0.2250\pm 0.0027$~\cite{Cabibbo:2003ea} is obtained
from four hyperon beta decays, where the quoted uncertainty is only experimental.
The central value is consistent with the one from kaon decays.
Contributions due to second order SU(3) breaking are not taken into account.

\subsection{\boldmath\Vub}
\begin{figure*}[t]
\centering
\includegraphics[width=80mm]{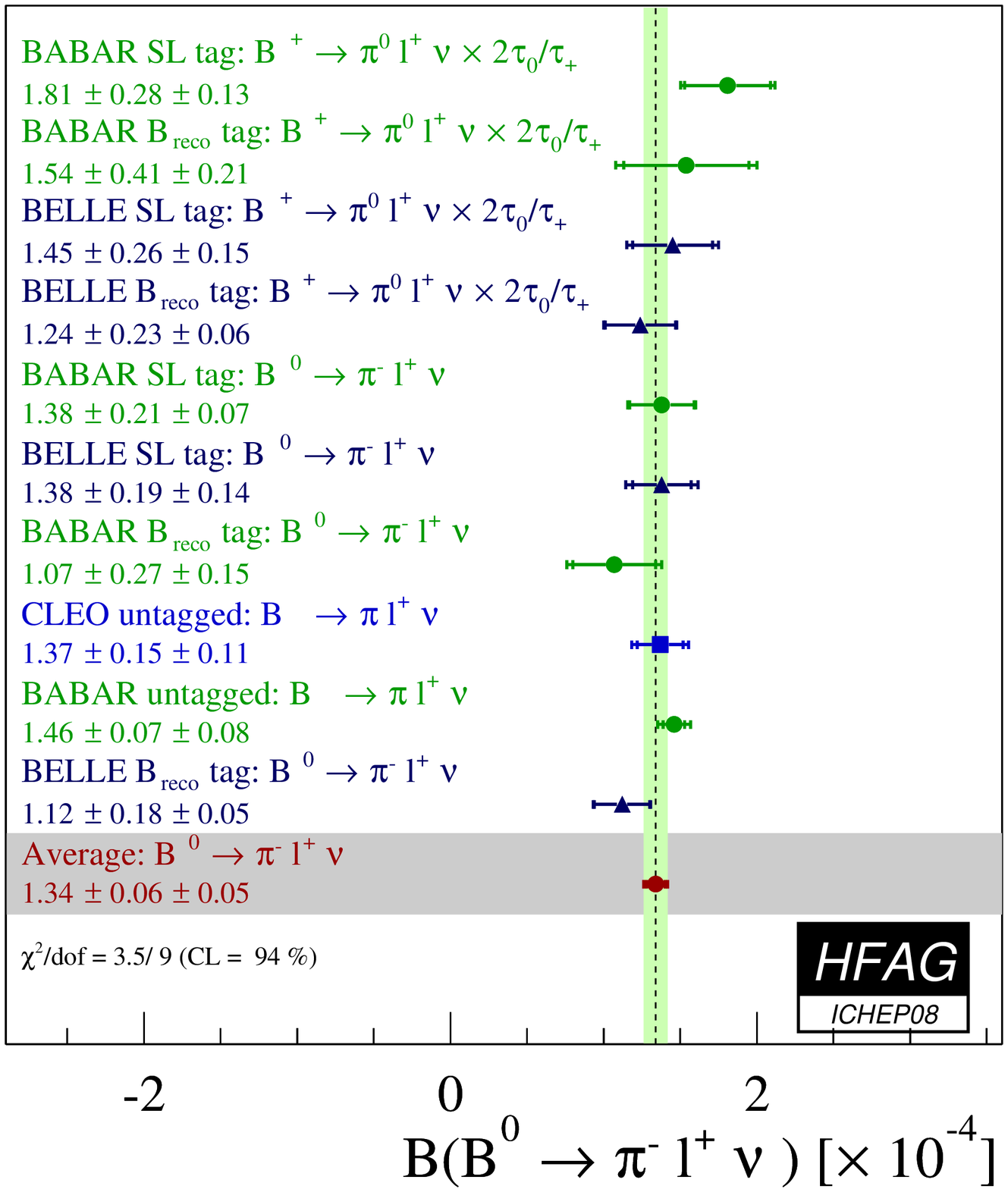}
\includegraphics[width=80mm]{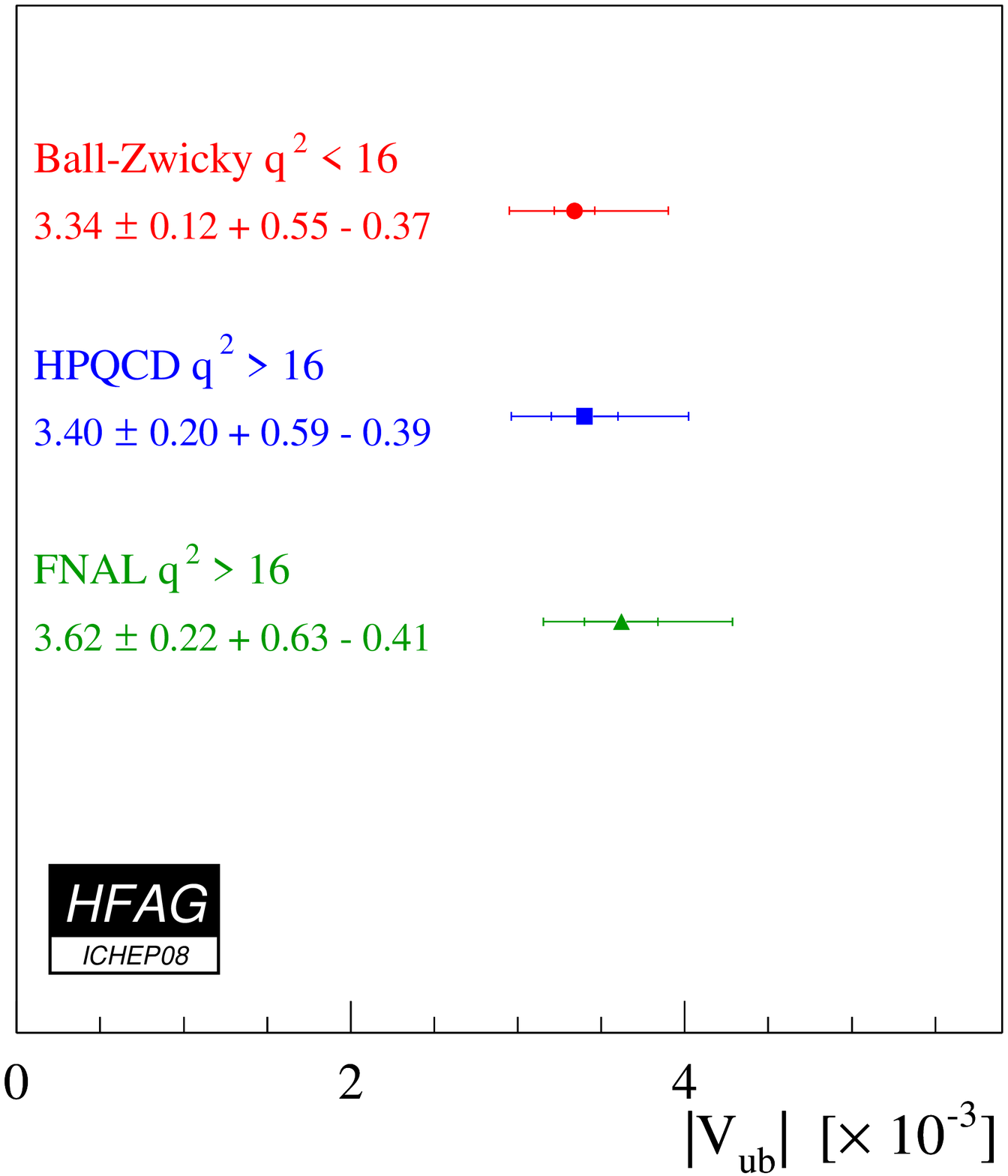}
\caption{$B\to \pi\ell\nu$ measured branching fractions and their average~\cite{hfag08} (left plot) and extracted value of \Vub~\cite{hfag08} (right plot) for different computations: Ball--Zwicky~\cite{Ball:2004ye}, which stands for the QCD light cone sum rule, HPQCD~\cite{Dalgic:2006dt} and FNAL~\cite{Okamoto:2004xg}, which stands for Fermilab/MILC.} \label{fig:vubexl}
\end{figure*}

Exclusive and inclusive semileptonic decays (\eg\ $B\rightarrow \pi\ell\nu$
and $B\rightarrow X_u\ell\nu$, where $X_u$ indicates the fragmentation products from the $u$ quark,
respectively) rely on different experimental and theoretical approaches,
thus providing a complementary way to extract \Vub. \\
\Vub\ can be extracted from exclusive charmless semileptonic decays, 
$B\rightarrow \pi\ell\nu$, where the corresponding rate is related to \Vub\ 
by the form factor $f_+(q^2)$, where $q^2$ is the momentum transfer squared to the lepton pair.
Non perturbative methods for the calculation of the form factors include
unquenched lattice QCD, where we use the HPQCD~\cite{Dalgic:2006dt} and 
Fermilab/MILC~\cite{Okamoto:2004xg}
calculations, which differ for the treatment of the $b$ quark, 
and QCD light cone sum rules~\cite{Ball:2004ye}.
Consistent results are obtained using earlier quenched QCD calculations~\cite{Abada:2000ty}.
Measurements of the $B\to \pi\ell\nu$ decays have been perfomed
by CLEO, \babar\ and \belle, exploiting different analysis techniques, where results 
are presented for the full $q^2$, $q^2>16$~GeV and $q^2<16$~GeV ranges. The last
two phase space regions correspond to regions where the lattice and QCD light cone sum rule 
calculations of the form factors are restricted to, respectively. 
The measurement techniques fall into two broad classes: untagged and tagged, 
depending on whether the $B$ in the event that does not decay into the
$\pi\ell\nu$ final state is tagged or not. Higher
statistics and higher background discriminate the first method from the second.
The corresponding measurements of the total branching ratio for all the collaborations 
and their average is shown in Fig.~\ref{fig:vubexl} (left plot). From the average, and using
both lattice QCD and QCD light cone sum rules, the value of 
\Vub\ is shown in Fig.~\ref{fig:vubexl} (right plot). The \Vub\ results 
coming from different theoretical calculations are consistent among themselves.
The dominant systematic in the \Vub\ extraction is due to the theoretical error. 
An improved treatment of the QCD light cone sum rule calculation was recently 
presented~\cite{Duplancic:2008tk, Duplancic:2008ix},
eventually giving consistent results for the mean value and the error on \Vub.
Concerning lattice QCD, an effort is underway to
perform a simultaneous fit to the experimental and lattice data using 
the model indipendent $z$ parametrization for the form factor~\cite{Lattice08:RuthVanDerWater}.
In particular, a 12 bin $q^2$ spectrum was measured by \babar~\cite{Aubert:2006px}.
More high precision $q^2$ measured spectra are foreseen in the future.

Moreover, still regarding exclusive semileptonic decays, 
experimental measurements of the $B\to \rho\ell\nu$ branching ratio have been
performed by \babar, \belle\ and CLEO and will provide a test of 
the \Vub\ extraction from $B\to \pi\ell\nu$
decays, once the corresponding form factors are computed.

\begin{figure*}[t]
\centering
\includegraphics[width=80mm]{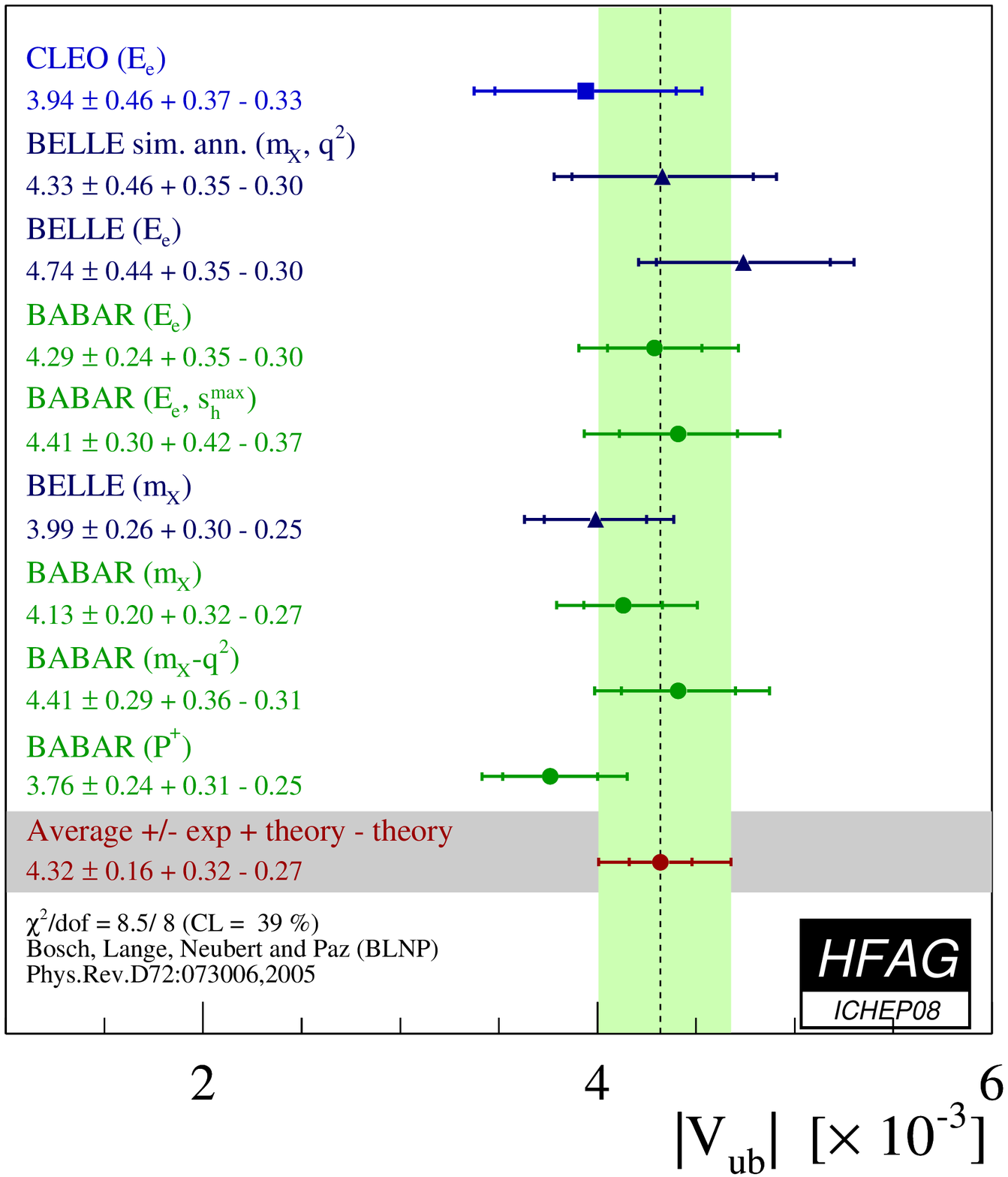}
\includegraphics[width=80mm]{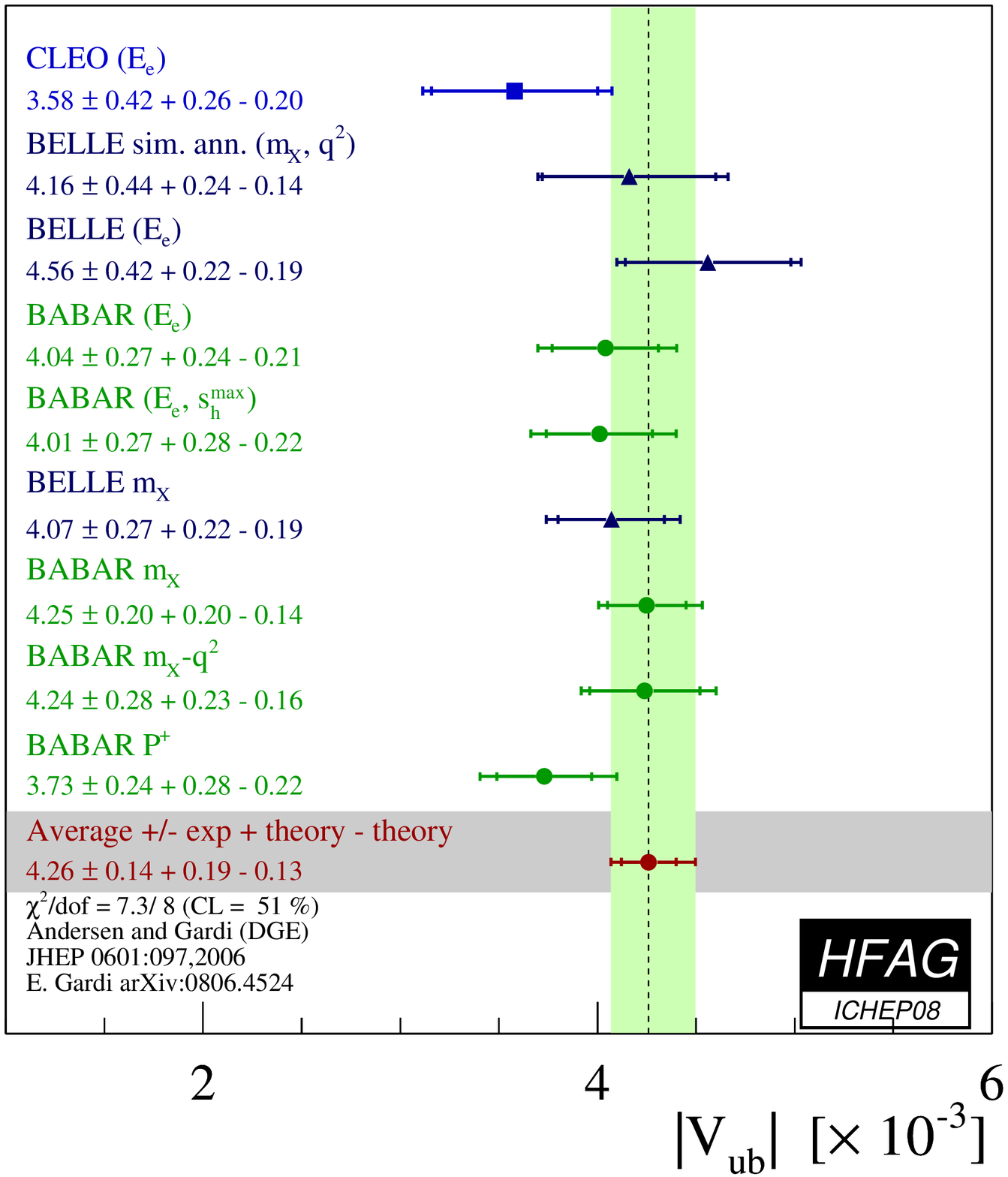}
\includegraphics[width=80mm]{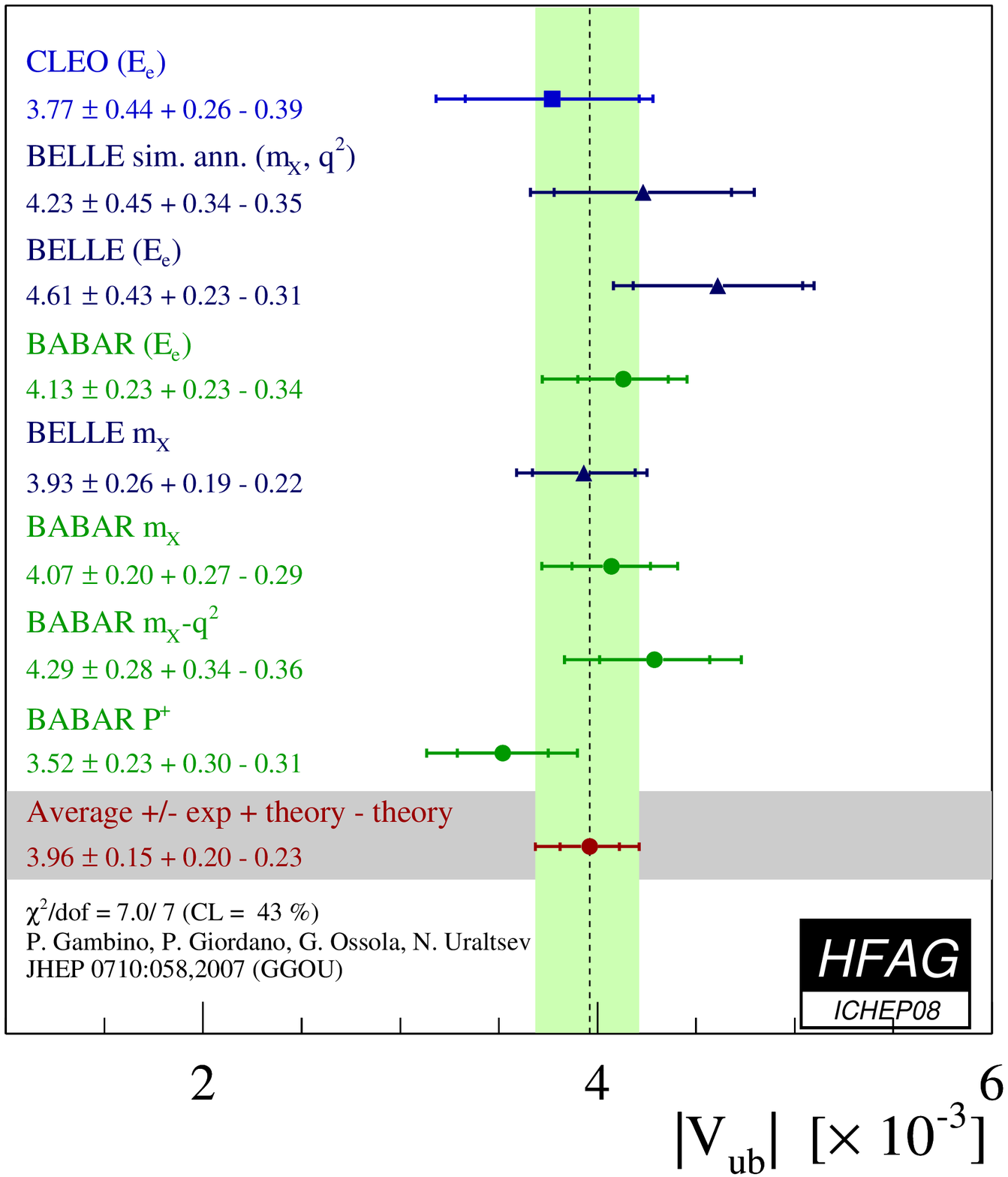}
\includegraphics[width=80mm]{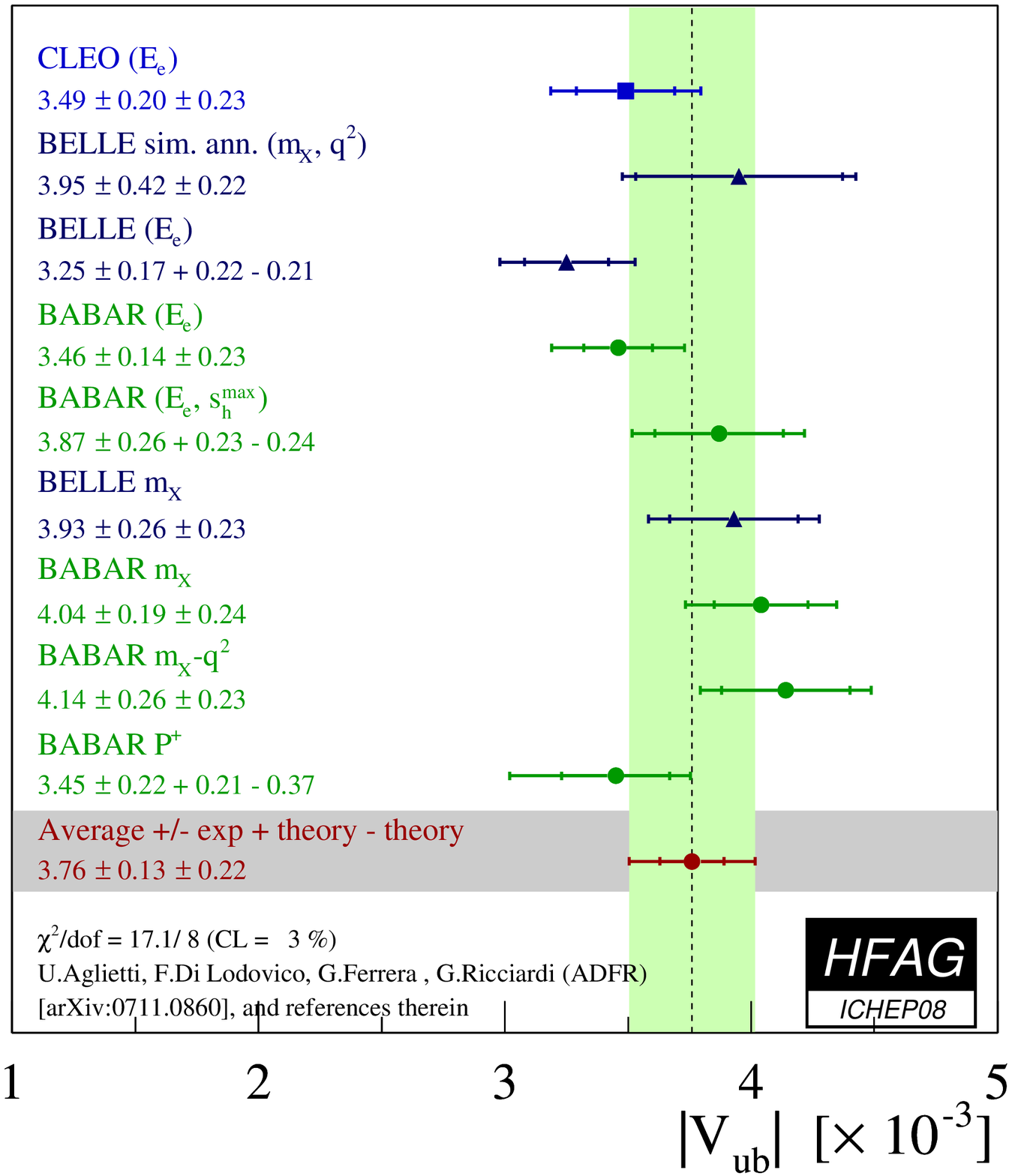}
\caption{Measured values of \Vub\ for the different analyses and theoretical
approaches~\cite{hfag08}.} \label{fig:vubincl}
\end{figure*}

The measurement of the inclusive decays rate for $B\to X_u \ell\nu$ decays is affected by
a large background of the order $\mathrm{{|V_{ub}/V_{cb}|}^2\sim 1/50}$, due to the
look--alike $B\to X_c \ell\nu$ decays. To suppress this background stringent kinematic
cuts are applied. Thus, a partial branching fraction, {\it i.e.} limited to the particular kinematic region
selected, which ranges from $\sim 20\%$ to $\sim 60\%$ of the total rate,
is measured. This challenges theory.
Whilst the total branching fraction can be computed using 
Heavy Flavour Expansion (HQE) and QCD perturbation theory, the partial rate needs
further theoretical tools, which have been the subject of intense theoretical effort, expecially in 
the last years.
The kinematic cuts are applied using the following variables:
the lepton energy (\el), the invariant mass of
the hadron final state (\mx), the light--cone distribution 
(\pplus $\equiv E_X - |\vec{p}_X|$, $E_X$ and $\vec{p}_X$ being the energy
and the magnitude of the 3--momentum of the hadronic system) and
a two dimensional distribution in the
electron energy and \smax,  the maximal \mx$^2$ at fixed $q^2$ and
\el. The differential rate needed from theory to extract \Vub\ from the experimental
results has been calculated using several different theoretical approaches. 
In chronological order, they are BLNP~\cite{Lange:2005yw} (a shape function approach, 
where the shape function represents the momentum distribution function of the 
$b$ quark in the $B$ meson), DGE~\cite{Andersen:2005mj, Gardi:2008bb}, 
(a resummation based approach), GGOU~\cite{Gambino:2007rp} (an HQE based structure function parametrization
approach) and ADFR~\cite{Aglietti:2007ik,Aglietti:2006yb}
(a soft gluon resummation and analytic time--like QCD coupling approach).
Concerning BLNP, recent NNLO corrections~\cite{Asatrian:2008uk} were presented. 
The models depend strongly on the $b$ quark mass, except for ADFR,
so it is very important to use a precise determination of the $b$ quark mass. 
The fit performed to obtain
the value of the $b$ quark mass is described in section~\ref{section:Vcb}.
The same value of the mass is used for the four models for consistency, translated to the
different mass schemes as needed~\cite{hfag08}.
The results obtained by these methods and the corresponding averages are shown in 
Fig.~\ref{fig:vubincl}. 

In inclusive decays, higher values of \Vub\ than from the exclusive decays are found,
although consistent within 1$\sigma$ from the Fermilab/MILC result.
The error on \Vub\ from inclusive decays is smaller than from exclusive decays.
The predicted value of \Vub\ from $sin2\beta$ favours values
closer to the exclusive one~\cite{Bona:2006ah}. Intense theoretical and experimental activity
is undergoing to compare the different methods among themselves and with the experimental 
results~\cite{CKM08:PaoloGambino}.
A slightly higher value of \Vub\ (\Vub $= (4.87 \pm 0.24 ^{+0.38}_{-0.38})\times 10^{-3}$) 
is obtained by another approach (BLL \cite{Bauer:2001rc}), OPE based, 
where a combined cut in the $(M_X, q^2)$ plane is proposed to reduce the theoretical
uncertainties. 
Moreover, a different strategy also has been adopted to overcome the problem of the
knowledge of the shape function. As the leading shape function 
can be measured in $B\to X_s \gamma$ decays, there
are prescriptions that relate directly the partial rates for $B\to X_s \gamma$
and $B\to X_u \ell\nu$ decays~\cite{Neubert:1993um,Leibovich:1999xf, 
Lange:2005qn,Lange:2005xz},
thus avoiding any parametrization of the SF.
However, uncertainties due to the sub--leading SF remain. Results with this method
have been obtained by \babar~\cite{Aubert:2006qi}
\Vub $= (4.92 \pm 0.32 \pm 0.36)\times 10^{-3}$, and Ref.~\cite{Golubev:2007cs}, 
using the \babar\ results in Ref.~\cite{Aubert:2005mg},
\Vub $= (4.28 \pm 0.29 \pm 0.29 \pm 0.26 \pm 0.28)\times 10^{-3}$,
\Vub $= (4.40 \pm 0.30 \pm 0.41 \pm 0.23)\times 10^{-3}$,
using Refs.\cite{Leibovich:1999xf} and \cite{Lange:2005qn,Lange:2005xz}, respectively.

There are several determinations of \Vub. Finally, we choose to quote as result for the inclusive 
decays the one from ADFR \Vub $= (3.76 \pm 0.13 \pm 0.22)\times 10^{-3}$.

Very recently, a preliminary result from \belle\ using a multivariate 
analysis~\cite{CKM08:PhillipUrquijo}, 
in which $\sim 90\%$ of the total rate is measured, has been presented.
This experimental measurement is
extremely interesting as it will help in a further understanding of \Vub\ from inclusive decays.

\section{SECOND ROW OF THE CKM MATRIX}
\subsection{\boldmath\Vcd}
The most precise measurement of the \Vcd\ matrix element comes from
neutrino production of charm at high energy, where the underline process is
a neutrino interacting with a $d$ quark, producing a charm quark that
fragments into a charmed hadron. The extracted value of \Vcd\ is
\Vcd $ = 0.230\pm 0.011$~\cite{Amsler:2008pdg} from the average
of several experiments. The dominant error comes from the mean
semi--muonic branching ratio for charmed hadrons produced in neutrino
anti--neutrino scattering, followed by the QCD scale uncertainty.

The CKM matrix element \Vcd\ can be determined also through the study of
semileptonic decays of the $D$ meson, where semileptonic decays are a preferred
way to determine the matrix elements as strong interaction effects are confined 
to the hadronic current. To extract the CKM matrix element from the
semileptonic decay rate the form factor, which measures the probability to
form the final state hadron, has to be input. The form factor is a function of
the $q^2$, the square of the transfer momentum to the lepton neutrino pair.
There are a variety of model dependent calculations of the form factor.
Here we will adopt the lattice QCD results~\cite{Aubin:2004ej}.
Decay rates have been measured by \belle~\cite{Widhalm:2006wz} 
and CLEO-c~\cite{:2007sm}.
Their average leads to the value \Vcd $ = 0.218\pm 0.007 \pm 0.023$,
where the errors are experimental and theoretical, respectively. 
The theory error is dominant.
Very recently, CLEO-c has published a new tagged analysis~\cite{Ge:2008yi},
which is partially overlapping with the untagged analysis in Ref.~\cite{:2007sm},
and it presents a consistent value of \Vcd. 

\subsection{\boldmath\Vcs}

The most precise measurement of the CKM matrix element \Vcs\ comes from
the study of semileptonic $D$ decays. The average of the
\babar~\cite{Aubert:2007wg}, \belle~\cite{Widhalm:2006wz} 
and CLEO-c~\cite{:2007sm} results is
\Vcs $ = 0.99 \pm 0.01 \pm 0.10$, where the errors are experimental
and theoretical, respectively, and the form factors from Ref~\cite{Aubin:2004ej}.
are adopted. Results from the CLEO-c tagged analysis~ \cite{Ge:2008yi}, very recently published,
are consistent with the one reported in Ref.~\cite{:2007sm}.

The leptonic $D_s$ decays, since no hadronic interactions are present
in the leptonic final state $\ell\nu$,
provide a very clean enviroment to determine \Vcs.
In these decays, strong interaction effects can be parametrized 
by the pseudoscalar decay constant $f_{D_s}$ which describes
the amplitude for the $c$ and $\overline{d}$ quarks within
the $D_s^+$ to have zero separation, a condition necessary to
annihilate into the virtual $W^+$ boson that produces the 
$\ell\nu $ pair.
Results from \babar~\cite{Aubert:2006sd}, 
\belle~\cite{:2007ws} CLEO-c~\cite{Artuso:2007zg,:2007zm},
assuming $f_{D_s}$ computed by lattice QCD~\cite{Follana:2007uv} give:
\Vcs $ = 1.07\pm 0.08$~\cite{Amsler:2008pdg}, 
where the error is dominated by the determination of $f_{D_s}$. This is the best
measurement of \Vcs, without assuming unitarity of the CKM matrix.

Decays of the $W$ boson at LEP2 have been used to determine \Vcs\ by Delphi,
tagging the decay $W\rightarrow cs$. The result is 
\Vcs $ = 0.94^{+0.32}_{-0.26} \pm 0.13$~\cite{Abreu:1998ap}, where the errors are statistical
and systematic, respectively. 
Moreover, using hadronic $W$ decay and assuming
the unitarity of the CKM matrix, from averages of the LEP
experiments, \Vcs $ = 0.9777\pm 0.014$~\cite{LEP:2005W} in obtained, where the total
error depends on the other CKM matrix elements, but it is
dominated by the experimental uncertainty.

The determination of \Vcs\ from neutrino and antineutrino interactions
suffers from the uncertainty of the $s$--quark sea content
leading to \Vcs\ $> 0.59$~\cite{Groom:2000pdg}.

\subsection{\boldmath\Vcb}
\label{section:Vcb}
\begin{figure*}[t]
\centering
\includegraphics[width=80mm]{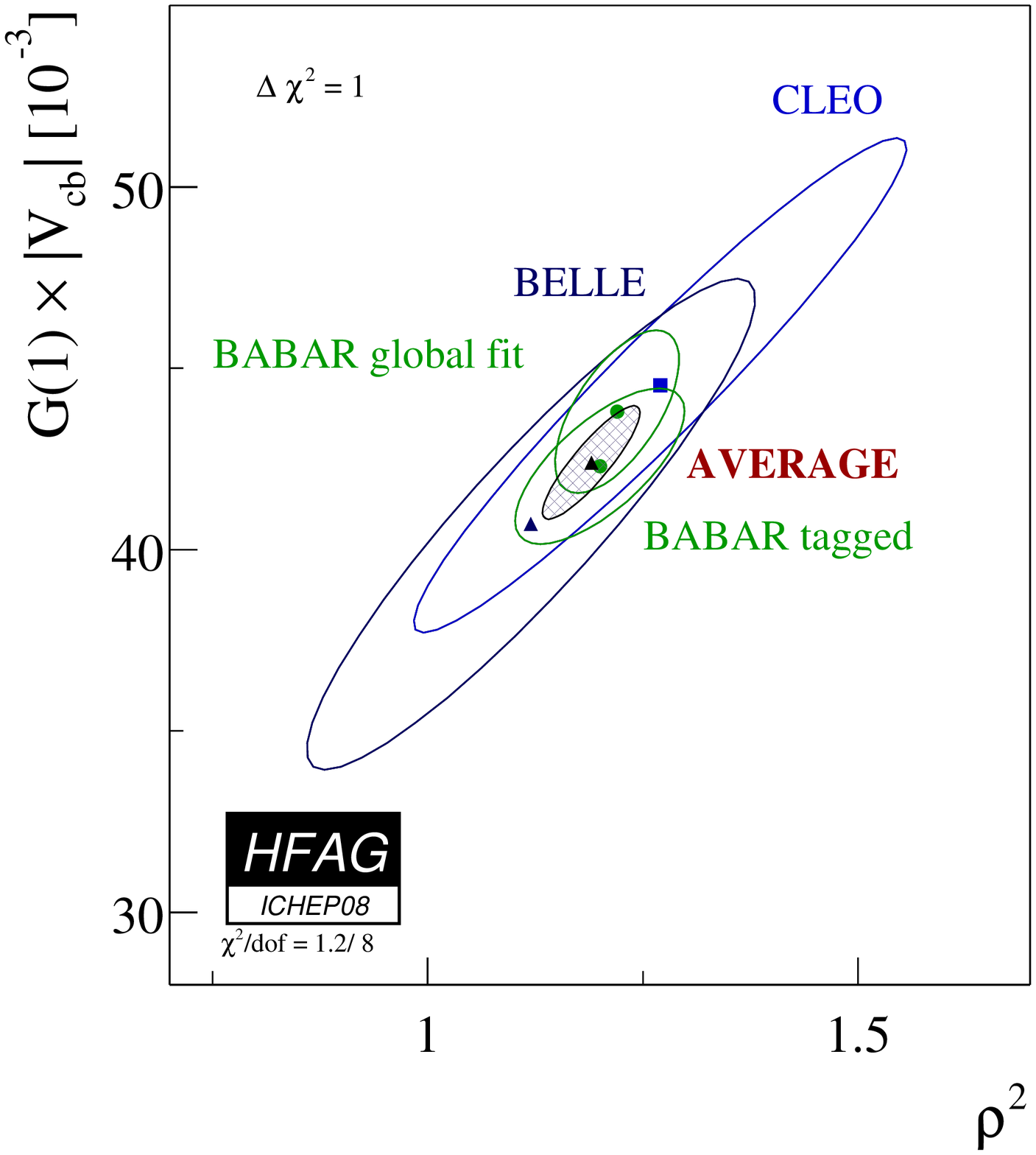}
\includegraphics[width=80mm]{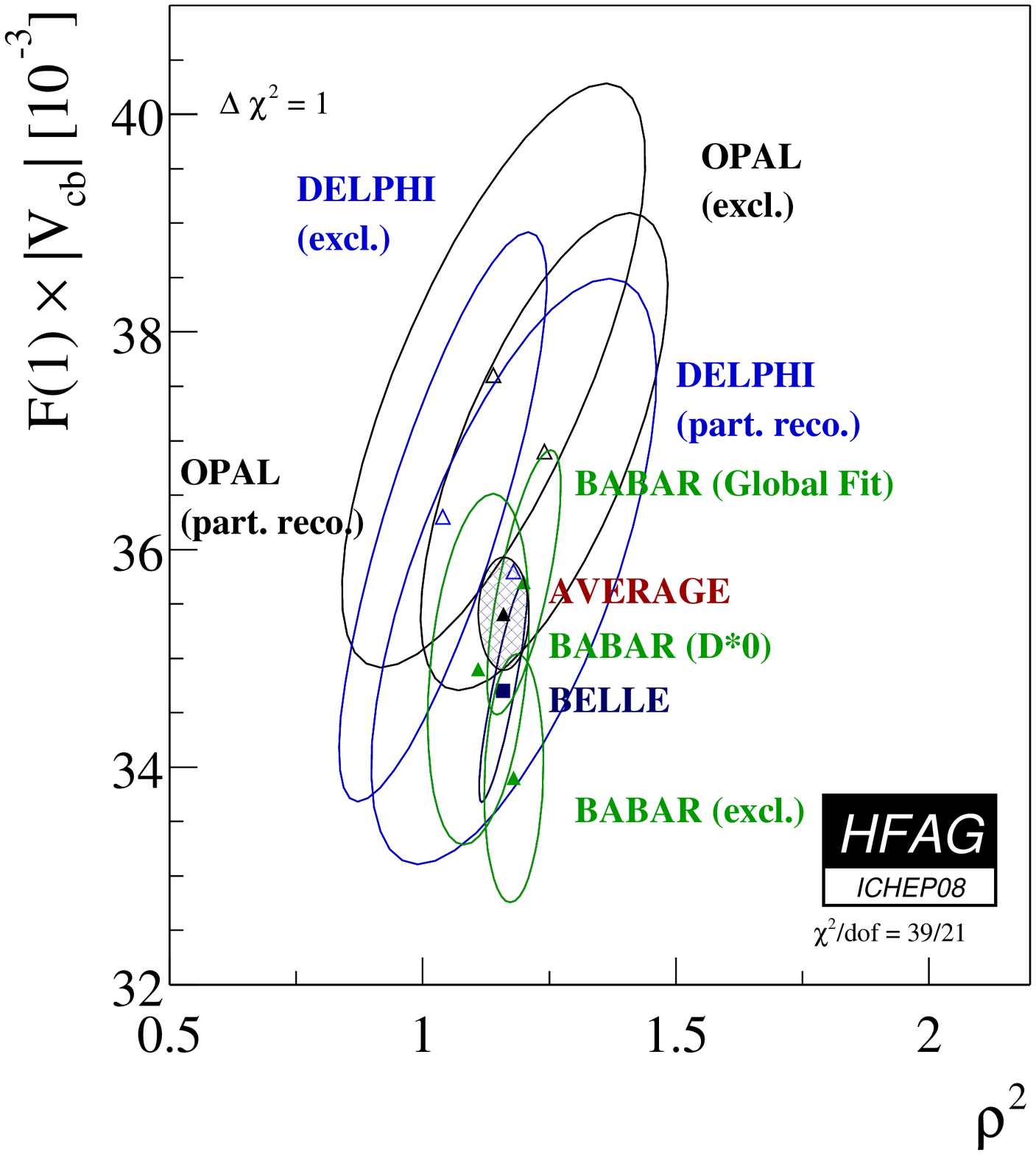}
\caption{Zoomed bi--dimensional plot of $G(1)$\Vcb~\cite{hfag08} (left plot) and $F(1)$\Vcb~\cite{hfag08} (right plot) versus the slope $\rho^2$.} \label{fig:vcbexl}
\end{figure*}

The \Vcb\ matrix element is determined from semileptonic
exclusive and inclusive $b\rightarrow c\ell \nu$ decays,
which rely on different theoretical calculations.
Several results were presented recently by \babar\ and \belle.
The determination of \Vcb\ from exclusive $b\rightarrow c\ell \nu$ decays
is based on the $B \rightarrow D^{(*)} \ell \nu$ decays, for which,
in the assumption of infinite $b$ and $c$ quark masses, the form
factors describing the $B\rightarrow D^{(*)}$ transitions depend
only on the product, $w$, of the initial, $v$, and final, $v^\prime$, state
hadron four--velocities, $w\equiv v\times v^\prime$, and relies on a parametrization
of the form factors using the Heavy Quark Symmetry (HQS)~\cite{Isgur:1989ed,
Isgur:1989vq,Shifman:1987rj} and a 
non--perturbative calculation of the form factor normalization at
$w=1$, which corresponds to the maximum momentum transfer to the leptons.
We adopt the parametrization from Ref.~\cite{Caprini:1997mu}, and lattice QCD
to correct the normalization of the form factor at $w=1$, due to the
finite quark masses. Experimentally, the $w$ spectrum is measured and
\Vcb\ is obtained from an extrapolation of the measured $w$ spectrum to 1.
Several analyses from \babar~\cite{Aubert:2007qs,Aubert:2007qw, 
Aubert:2008yv,:2008ii}, and \belle~\cite{Adachi:2008nd}, 
which adopt different experimental techniques, were recently presented.
In particular, in Ref.~\cite{Aubert:2008yv} $B \rightarrow D^{(*)} \ell \nu$ are 
selected applying for the first time a global fit to $D^{0(+)}\ell$ 
reconstructed combinations in a three dimensional space of kinematic 
variables to determine their branching fractions and the form factor parameters.
The form factors for the $B \rightarrow D \ell \nu$ and
$B \rightarrow D^{*} \ell \nu$ decays are $G(w)$ and $F(w)$, respectively.
The bi--dimensional plots of the form factor at $w=1$ times \Vcb\ versus the slope
parameter for the form factors $\rho$ is shown in Fig.~\ref{fig:vcbexl}, whose
fitted values are $G(1)$\Vcb $=(42.4\pm 1.6)\times 10^{-3}$, $\rho^2=1.16\pm0.05$ 
and $F(1)$\Vcb $=(35.41\pm0.52)\times 10^{-3}$, $\rho^2=1.19\pm0.06$ , 
where the two slopes are two different parameters.
Assuming $G(1)=1.074\pm0.018\pm0.016$~\cite{Okamoto:2004xg} and
$F(1)=0.924\pm 0.012 \pm 0.019$~\cite{Bernard:2008dn}, where
the errors are statistical and systematical, respectively, 
once corrected by a factor 1.007 for QCD effects,  
\Vcb $=(39.2\pm 1.5 \pm 0.9)\times 10^{-3}$
and \Vcb $=(38.2\pm 0.6 \pm 1.0)\times 10^{-3}$~\cite{hfag08}
are obtained, for $B \rightarrow D \ell \nu$ and
$B \rightarrow D^{*} \ell \nu$ decays, respectively, where the errors
are $\sim 10\%$, which means an improvement of $\sim 50\%$ with respect to the 
previous results mainly thanks to Ref.~\cite{Aubert:2008yv}, 
and $\sim 3-4\%$. The two results are completely consistent.

Inclusive $b\rightarrow c \ell \nu$ decays, $B\rightarrow X_c\ell\nu$,
where $X_c$ indicates the fragmentation products from the $c$ quark, 
can be used to determine \Vcb\ from their branching fraction
and with parameters that describe the motion of the $b$ quark in the
$B$ meson. These parameters, within the framework of the 
Heavy Quark Expansion (HQE), include the $b$ quark mass, $m_b$.
\Vcb, $m_b$ and other parameters can be extracted simultaneously
from a global fit to the measured moment of the leptonic energy and hadronic
mass spectra. Moreover, also moments from $b\rightarrow s \gamma$
decays are included in the global fit, as at Leading Order they can be
represented by the same shape function. Higher order corrections
are expected to be small~\cite{GardiGambino}. 
To note however that just a fit to the $b\rightarrow s \gamma$ moments tends
to give a value of $m_b$ about 1$\sigma$ lower than the one from
$b\rightarrow s \gamma$ and $b\rightarrow c \ell \nu$ moments combined.
However, the latest is consistent with the measurement coming from the study of
bottonium resonances~\cite{Kuhn:2007vp}.
The fitted value of \Vcb, using expressions in the kinetic 
scheme~\cite{Benson:2003kp,Gambino:2004qm}, is: 
\Vcb $=(41.67\pm 0.43 \pm 0.08 \pm 0.58)\times 10^{-3}$~\cite{hfag08}, 
where the errors are due to the global fit, the $B$ lifetime and theory, respectively.
This value is more than 2$\sigma$ higher than the corresponding values from
the $B \rightarrow D^{(*)} \ell \nu$ decays, but twice more precise, expecially
with respect to the $B \rightarrow D \ell \nu$ decays, dominated by the 
experimental uncertainty. 
Whilst $B \rightarrow D \ell \nu$ decays account for $\sim 70\%$ of the 
total $b\rightarrow c \ell \nu$ rate, the contribution of resonant and
non resonant decays to other charm states is not very well measured
and may help to explain the difference in the \Vcb\ 
determination~\cite{Pegna:2008uw}, 
whereas the current difference between the total exclusive and inclusive
final states is $\sim 10\%$.
%
%

\section{THIRD ROW OF THE CKM MATRIX}
\subsection{{\boldmath\Vtd} AND {\boldmath\Vts}}

The top quark is expected to decay almost completely
to a $W$ boson and a $b$ quark (the corresponding rate is 
$>99.8\%$ at 90\%\ CL), so the CKM matrix elements 
\Vtd\ and \Vts\ are measured indirectly from the 
$B$--$\overline{B}$ oscillations mediated by box diagrams
with top quarks, or from the loop--mediated rate of $K$ and $B$ decays.
The major uncertainty in the extraction of the parameters 
comes from the theoretical hadronic uncertainties.
The time--integrated measurements of $B^0\overline{B^0}$ mixing
have been performed for both the $B_d$ and more recently also the
$B_s$ mesons.
The measured mass differences $\Delta{m_d}$ 
for the neutral $B_d$ meson mass eigenstates
was measured by many collaborations, using a 
variety of different techniques. A high precision was achieved,
and the corresponding average~\cite{Amsler:2008pdg}, assuming
a decay width difference $\Delta{\Gamma_d}=0$ and no $CP$ violation in mixing, 
is $\Delta{m_d}=(0.507\pm 0.005)$ps$^{-1}$,
dominated by the $B$--factories \babar\ and \belle. 
The statistical and systematic errors equally 
contribute to the final uncertainty. The squared mass difference
$\Delta{m_d}^2$ is related to \Vtd\ through the product of the
$B_d$ decay constant and the bag factor: $f_{B_d}\sqrt{{B}_{B_d}}$. 
Using the unquenched lattice QCD calculation, 
$f_{B_d}\sqrt{{B}_{B_d}}= (225\pm 25)$~MeV~\cite{Lubicz:2008am}, the value
\Vtd $= (8.0\pm 0.9)\times 10^{-3}$ is obtained, whereas the theory uncertainty
dominates.\\
The $B_s^0\overline{B_s^0}$ oscillation has been observed for the first
time in 2006 by CDF~\cite{Abulencia:2006ze}. 
The measured mass difference is $\Delta{m_s}=(17.77\pm 0.10\pm 0.07)$ps$^{-1}$,
where the errors are statistical and systematic, respectively.
Similarly, there are studies by D0~\cite{D0Coll:Bsmixing}, which also
hint to oscillations in $B_s$ mixing at $\sim 2\sigma$ level. 
Using the measured value of $\Delta{m_s}$, and the product
$f_{B_s}\sqrt{{B}_{B_s}}= (270\pm 30)$~MeV~\cite{Lubicz:2008am} from
unquenched lattice QCD calculations, the value
\Vtd $= (39.4\pm 4.4)\times 10^{-3}$ is obtained, where again 
the dominant uncertainty is due the lattice QCD. 

However,
if the ratio $\Delta{m_d}/\Delta{m_s}$ is calculated, the corresponding
theoretical uncertainties decrease. 
Being $f_{B_s}\sqrt{{B}_{B_s}}/f_{B_d}\sqrt{{B}_{B_d}}=1.21\pm\pm0.04$~\cite{Lubicz:2008am},
the value of the ratio of the two mass differences is:
\Vtd$/$\Vts $= (0.206\pm 0.001\pm 0.007)\times 10^{-3}$, whose error
is greatly reduced with respect to the individual computations.

The radiative penguins decays $b\rightarrow d \gamma$ and $b\rightarrow s\gamma$
constitute an independent way than $B$ oscillation to measure the matrix 
elements \Vtd\ and \Vts, respectively, as they are affected by different 
experimental and theoretical (there are penguin instead of box diagrams) uncertainties. 
Recent results were presented by both \babar\ and \belle\ for
exclusive~\cite{:2008cy,:2008gf,Taniguchi:2008ty} and inclusive~\cite{:2008gvb} decays.
The extracted values of \Vtd$/$\Vts are \Vtd$/$\Vts $= (0.233\pm 0.025\pm 0.022)\times 10^{-3}$ and \Vtd$/$\Vts $= (0.195\pm 0.020\pm 0.015)\times 10^{-3}$ for the 
\babar\ and \belle\ exclusive analyses, respectively, and
\Vtd$/$\Vts $= (0.177\pm 0.043\pm 0.001)\times 10^{-3}$ for the \babar\ inclusive 
analysis.
They are consistent with the result from $B$ oscillations,
but with an error about 5$\sigma$ larger.

The measurement of the $K\to \pi\ell\nu$ branching fraction can lead to the cleanest determination
of $\mathrm{|V_{td}}\mathrm{V_{ts}^*|}$. Only three events have
been observed so far by E949~\cite{Anisimovsky:2004hr}, but a
statistically more accurate measurement is foreseen in the future by
NA62~\cite{Biino:2008ja}.

\subsection{\boldmath\Vtb}

All direct measurements of production and decay of the
top quark have been performed by the CDF and D0 collaborations at 
Fermilab.  
Measuring the ratio of the top quark branching fractions 
$R=\dfrac{({\cal B}t\to Wb)}{({\cal B}t\to Wq)}$, a value of \Vtb\ can
be extracted assuming unitarity 
$R=\dfrac{\mathrm{|V_{tb}|}^2}{\mathrm{|V_{tb}|}^2+\mathrm{|V_{ts}|}^2+\mathrm{|V_{td}|}^2}$. 
Both results from CDF~\cite{Acosta:2005hr} and D0~\cite{Abazov:2008yn} are available.
The limit is $R = 0.97 ^{+0.09}_{-0.08}$ with a total uncertainty of
about 9$\%$, where the uncertainty is statistical and systematic.
The largest uncertainty comes with the limited statistics.
The corresponding limit on \Vtb\ is \Vtb$>0.89$ at 95$\%$ CL.

Single top quark events can be used to study the $Wtb$ coupling and thus,
without any unitarity assumption, directly extracting \Vtb. The first
observation of the single top production is presented by 
D0~\cite{Abazov:2006gd}. Similarly, evidence for single top productions
is found in the CDF data~\cite{Aaltonen:2008sy}. Higher statistics was used,
but a lower cross section is indicated. D0 measures a single top cross
section $\sigma = (4.9 \pm 1.4 )$~pb,
where the uncertainty is statistical and systematic. 
The largest uncertainty comes with the limited statistics.
The extracted value of \Vub\ is \Vub = $(1.3 \pm 0.2)$.

\section{SUMMARY} 
A review of the current status of the CKM matrix elements is given, with particular
attention to the latest results. For each matrix element, there are different
processes which can be used to determine its value. Choosing the determination with the
smallest error, the CKM matrix is:
\begin{equation}
\begin{pmatrix}
|V_{ud}| &|V_{us}| &|V_{ub}| \\
|V_{cd}| &|V_{cs}| &|V_{cb}| \\ 
\multicolumn{2}{c}{|V_{td}|/|V_{ts}|}& |V_{tb}| \\
\end{pmatrix}
=
\begin{pmatrix}
0.97408(26) & 0.2246(12) & 0.00376(26) \\
0.230(11)   & 0.99(10)   & 0.04167(72)   \\
\multicolumn{2}{c}{0.000206(7)} & 1.3(2)\\
\end{pmatrix}
\end{equation}

where only the total error per each matrix element is shown.
A significant progress has been made in the past years (\eg\ understanding of the \Vus\ value from kaon decays,
more \Vub\ theoretical calculations, etc.), and more work is needed in the future
to achieve a better understand of the current measurements as highlighted in the text. 

\begin{acknowledgments}
We are grateful to the experimental collegues from \babar, \belle, CDF, CLEO and
D0 for help in the preparation of this talk, and having shared their latest results
when needed. 
Also, we would like to thank the theory collegues for several discussions
regarding the \Vub\ and \Vcb\ calculations. Finally, we 
we would like to thank the ICHEP08 organizers for the opportunity to present this talk 
and a flawless organization.
\end{acknowledgments}


\begin{thebibliography}{9}   

\bibitem{Towner:2007np}
  I.~S.~Towner and J.~C.~Hardy,
  Phys.\ Rev.\  C {\bf 77}, 025501 (2008)
  [arXiv:0710.3181 [nucl-th]].


\bibitem{Eronen:2007qc}
  T.~Eronen {\it et al.},
  Phys.\ Rev.\ Lett.\  {\bf 100}, 132502 (2008)
  [Erratum-ibid.\  {\bf 100}, 149902 (2008)]
  [arXiv:0712.3463 [nucl-ex]].

\bibitem{Amsler:2008pdg}
  C.~Amsler {\it et al.},Phys.\ Lett.\  B {\bf 667}, 1 (2008).

\bibitem{Serebrov:2004zf}
  A.~Serebrov {\it et al.},
  Phys.\ Lett.\  B {\bf 605}, 72 (2005)
  [arXiv:nucl-ex/0408009].


\bibitem{Pocanic:2003pf}
  D.~Pocanic {\it et al.},
  Phys.\ Rev.\ Lett.\  {\bf 93}, 181803 (2004)
  [arXiv:hep-ex/0312030].

\bibitem{Antonelli:2008jg}
  M.~Antonelli {\it et al.}  [FlaviaNet Working Group on Kaon Decays],
  arXiv:0801.1817 [hep-ph].

\bibitem{Lai:2007dx}
  A.~Lai {\it et al.}  [NA48 Collaboration],
  Phys.\ Lett.\  B {\bf 647}, 341 (2007)
  [arXiv:hep-ex/0703002].

\bibitem{Antonio:2007mh}
  D.~J.~Antonio {\it et al.},
  arXiv:hep-lat/0702026.

\bibitem{Follana:2007uv}
  E.~Follana, C.~T.~H.~Davies, G.~P.~Lepage and J.~Shigemitsu  [HPQCD
                  Collaboration and UKQCD Collaboration],
  Phys.\ Rev.\ Lett.\  {\bf 100}, 062002 (2008)
  [arXiv:0706.1726 [hep-lat]].

\bibitem{Le Diberder:1992fr}
  F.~Le Diberder and A.~Pich,
  Phys.\ Lett.\  B {\bf 289}, 165 (1992).

\bibitem{Banerjee:2008hg}
  S.~Banerjee and f.~t.~B.~Collaboration,
  arXiv:0811.1429 [hep-ex].

\bibitem{Maltman:2008ib}
  K.~Maltman, C.~E.~Wolfe, S.~Banerjee, J.~M.~Roney and I.~Nugent,
  arXiv:0807.3195 [hep-ph].


\bibitem{Ademollo:1964sr}
  M.~Ademollo and R.~Gatto,
  Phys.\ Rev.\ Lett.\  {\bf 13}, 264 (1964).

\bibitem{Cabibbo:2003ea}
  N.~Cabibbo, E.~C.~Swallow and R.~Winston,
  Phys.\ Rev.\ Lett.\  {\bf 92}, 251803 (2004)
  [arXiv:hep-ph/0307214].

\bibitem{Dalgic:2006dt}
  E.~Dalgic, A.~Gray, M.~Wingate, C.~T.~H.~Davies, G.~P.~Lepage and J.~Shigemitsu,
  Phys.\ Rev.\  D {\bf 73}, 074502 (2006)
  [Erratum-ibid.\  D {\bf 75}, 119906 (2007)]
  [arXiv:hep-lat/0601021].


\bibitem{Okamoto:2004xg}
  M.~Okamoto {\it et al.},
  Nucl.\ Phys.\ Proc.\ Suppl.\  {\bf 140}, 461 (2005)
  [arXiv:hep-lat/0409116].

\bibitem{Ball:2004ye}
  P.~Ball and R.~Zwicky,
  Phys.\ Rev.\  D {\bf 71}, 014015 (2005)
  [arXiv:hep-ph/0406232].

\bibitem{Abada:2000ty}
  A.~Abada, D.~Becirevic, P.~Boucaud, J.~P.~Leroy, V.~Lubicz and F.~Mescia,
  Nucl.\ Phys.\  B {\bf 619}, 565 (2001)
  [arXiv:hep-lat/0011065].

\bibitem{hfag08}
ICHEP08 updates available online at:
http://www.slac.stanford.edu/xorg/hfag/semi/ichep08/home.shtml

\bibitem{Duplancic:2008tk}
  G.~Duplancic and B.~Melic,
  Phys.\ Rev.\  D {\bf 78}, 054015 (2008)
  [arXiv:0805.4170 [hep-ph]].

\bibitem{Duplancic:2008ix}
  G.~Duplancic, A.~Khodjamirian, T.~Mannel, B.~Melic and N.~Offen,
  JHEP {\bf 0804}, 014 (2008)
  [arXiv:0801.1796 [hep-ph]].

\bibitem{Aubin:2004ej}
  C.~Aubin {\it et al.}  [Fermilab Lattice Collaboration and MILC
                  Collaboration and HPQCD Collab],
  Phys.\ Rev.\ Lett.\  {\bf 94}, 011601 (2005)
  [arXiv:hep-ph/0408306].

\bibitem{Widhalm:2006wz}
  L.~Widhalm {\it et al.},
  Phys.\ Rev.\ Lett.\  {\bf 97}, 061804 (2006)
  [arXiv:hep-ex/0604049].

\bibitem{Lattice08:RuthVanDerWater}
  R.~Van de Water, Lattice 2008, The XXVI International Symposium on Lattice Field Theory. Williamsburg, Virginia, USA, 2008 July 14–19. 

\bibitem{Aubert:2006px}
  B.~Aubert {\it et al.}  [BABAR Collaboration],
  Phys.\ Rev.\ Lett.\  {\bf 98}, 091801 (2007)
  [arXiv:hep-ex/0612020].

\bibitem{Lange:2005yw}
  B.~O.~Lange, M.~Neubert and G.~Paz,
  Phys.\ Rev.\  D {\bf 72}, 073006 (2005)
  [arXiv:hep-ph/0504071].

\bibitem{Asatrian:2008uk}
  H.~M.~Asatrian, C.~Greub and B.~D.~Pecjak,
  arXiv:0810.0987 [hep-ph].

\bibitem{Andersen:2005mj}
  J.~R.~Andersen and E.~Gardi,
  JHEP {\bf 0601}, 097 (2006)
  [arXiv:hep-ph/0509360].

\bibitem{Gardi:2008bb}
  E.~Gardi,
  arXiv:0806.4524 [hep-ph].

\bibitem{Gambino:2007rp}
  P.~Gambino, P.~Giordano, G.~Ossola and N.~Uraltsev,
  JHEP {\bf 0710}, 058 (2007)
  [arXiv:0707.2493 [hep-ph]].

\bibitem{Aglietti:2007ik}
  U.~Aglietti, F.~Di Lodovico, G.~Ferrera and G.~Ricciardi,
  accepted by EPJC [arXiv:0711.0860 [hep-ph]].

\bibitem{Aglietti:2006yb}
  U.~Aglietti, G.~Ferrera and G.~Ricciardi,
  Nucl.\ Phys.\  B {\bf 768}, 85 (2007)
  [arXiv:hep-ph/0608047].

\bibitem{Bona:2006ah}
  M.~Bona {\it et al.}  [UTfit Collaboration],
  JHEP {\bf 0610}, 081 (2006)
  [arXiv:hep-ph/0606167].


\bibitem{CKM08:PaoloGambino}
  P.~Gambino, PDG 2008, Workshop on the Unitarity Triangle, Rome, Italy, 2008 September 9-13.


\bibitem{Bauer:2001rc}
  C.~W.~Bauer, Z.~Ligeti and M.~E.~Luke,
  Phys.\ Rev.\  D {\bf 64}, 113004 (2001)
  [arXiv:hep-ph/0107074].

\bibitem{Neubert:1993um}
  M.~Neubert,
  Phys.\ Rev.\  D {\bf 49}, 4623 (1994)
  [arXiv:hep-ph/9312311].

\bibitem{Leibovich:1999xf}
  A.~K.~Leibovich, I.~Low and I.~Z.~Rothstein,
  Phys.\ Rev.\  D {\bf 61}, 053006 (2000)
  [arXiv:hep-ph/9909404].

\bibitem{Lange:2005qn}
  B.~O.~Lange, M.~Neubert and G.~Paz,
  JHEP {\bf 0510}, 084 (2005)
  [arXiv:hep-ph/0508178].

\bibitem{Lange:2005xz}
  B.~O.~Lange,
  JHEP {\bf 0601}, 104 (2006)
  [arXiv:hep-ph/0511098].

\bibitem{Aubert:2006qi}
  B.~Aubert {\it et al.}  [BABAR Collaboration],
  Phys.\ Rev.\ Lett.\  {\bf 96}, 221801 (2006)
  [arXiv:hep-ex/0601046].

\bibitem{Golubev:2007cs}
  V.~B.~Golubev, Y.~I.~Skovpen and V.~G.~Luth,
  Phys.\ Rev.\  D {\bf 76}, 114003 (2007)
  [arXiv:hep-ph/0702072].

\bibitem{Aubert:2005mg}
  B.~Aubert {\it et al.}  [BABAR Collaboration],
  Phys.\ Rev.\  D {\bf 73}, 012006 (2006)
  [arXiv:hep-ex/0509040].

\bibitem{CKM08:PhillipUrquijo}
  P.~Urquijo, PDG 2008, Workshop on the Unitarity Triangle, Rome, Italy, 2008 September 9-13.

\bibitem{:2007sm}
  S.~Dobbs {\it et al.}  [CLEO Collaboration],
  Phys.\ Rev.\  D {\bf 77}, 112005 (2008)
  [arXiv:0712.1020 [hep-ex]].

\bibitem{Ge:2008yi}
  J.~Y.~Ge {\it et al.} [CLEO Collaboration],
  arXiv:0810.3878 [hep-ex].

\bibitem{Groom:2000pdg}
  D.E.~Groom {\it et al.}, The European Physical Journal C15 (2000) 1.

\bibitem{Aubert:2007wg}
  B.~Aubert {\it et al.}  [BABAR Collaboration],
  arXiv:0704.0020 [hep-ex].

\bibitem{Aubert:2006sd}
  B.~Aubert {\it et al.}  [BABAR Collaboration],
  Phys.\ Rev.\ Lett.\  {\bf 98}, 141801 (2007)
  [arXiv:hep-ex/0607094].

\bibitem{:2007ws}
  K.~Abe {\it et al.}  [Belle Collaboration],
  Phys.\ Rev.\ Lett.\  {\bf 100}, 241801 (2008)
  [arXiv:0709.1340 [hep-ex]].

\bibitem{Artuso:2007zg}
  M.~Artuso {\it et al.}  [CLEO Collaboration],
  Phys.\ Rev.\ Lett.\  {\bf 99}, 071802 (2007)
  [arXiv:0704.0629 [hep-ex]].

\bibitem{:2007zm}
  K.~M.~Ecklund {\it et al.}  [CLEO Collaboration],
  Phys.\ Rev.\ Lett.\  {\bf 100}, 161801 (2008)
  [arXiv:0712.1175 [hep-ex]].

\bibitem{Abreu:1998ap}
  P.~Abreu {\it et al.}  [DELPHI Collaboration],
  Phys.\ Lett.\  B {\bf 439}, 209 (1998).

\bibitem{LEP:2005W}
LEP Collaborations, LEPWWG/XSEC/2005-1.

\bibitem{Isgur:1989ed}
  N.~Isgur and M.~B.~Wise,
  Phys.\ Lett.\  B {\bf 237}, 527 (1990).

\bibitem{Isgur:1989vq}
  N.~Isgur and M.~B.~Wise,
  Phys.\ Lett.\  B {\bf 232}, 113 (1989).

\bibitem{Shifman:1987rj}
  M.~A.~Shifman and M.~B.~Voloshin,
  Sov.\ J.\ Nucl.\ Phys.\  {\bf 47}, 511 (1988)
  [Yad.\ Fiz.\  {\bf 47}, 801 (1988)].


\bibitem{Caprini:1997mu}
  I.~Caprini, L.~Lellouch and M.~Neubert,
  Nucl.\ Phys.\  B {\bf 530}, 153 (1998)
  [arXiv:hep-ph/9712417].

\bibitem{Aubert:2007qs}
  B.~Aubert {\it et al.}  [BABAR Collaboration],
  Phys.\ Rev.\ Lett.\  {\bf 100}, 231803 (2008)
  [arXiv:0712.3493 [hep-ex]].


\bibitem{Aubert:2007qw}
  B.~Aubert {\it et al.}  [BABAR Collaboration],
  Phys.\ Rev.\ Lett.\  {\bf 100}, 151802 (2008)
  [arXiv:0712.3503 [hep-ex]].

\bibitem{Aubert:2008yv}
  B.~Aubert {\it et al.}  [BABAR Collaboration],
  arXiv:0809.0828 [hep-ex].

\bibitem{Aubert:2008yv}
  B.~Aubert {\it et al.}  [BABAR Collaboration],
  arXiv:0809.0828 [hep-ex].

\bibitem{:2008ii}
  B.~Aubert {\it et al.}  [BABAR Collaboration],
  arXiv:0807.4978 [hep-ex].

\bibitem{Adachi:2008nd}
  I.~Adachi {\it et al.} [The Belle Collaboration],
  arXiv:0810.1657 [hep-ex].

\bibitem{Bernard:2008dn}
  C.~Bernard {\it et al.},
  arXiv:0808.2519 [hep-lat].

\bibitem{Benson:2003kp}
  D.~Benson, I.~I.~Bigi, T.~Mannel and N.~Uraltsev,
  Nucl.\ Phys.\  B {\bf 665}, 367 (2003)
  [arXiv:hep-ph/0302262].

\bibitem{Gambino:2004qm}
  P.~Gambino and N.~Uraltsev,
  Eur.\ Phys.\ J.\  C {\bf 34}, 181 (2004)
  [arXiv:hep-ph/0401063].

\bibitem{GardiGambino}
Private communications with Paolo Gambino and Einan Gardi.

\bibitem{Kuhn:2007vp}
  J.~H.~Kuhn, M.~Steinhauser and C.~Sturm,
  Nucl.\ Phys.\  B {\bf 778}, 192 (2007)
  [arXiv:hep-ph/0702103].

\bibitem{Pegna:2008uw}
  D.~L.~Pegna and f.~t.~B.~collaboration,
  arXiv:0810.3706 [hep-ex].

\bibitem{Lubicz:2008am}
  V.~Lubicz and C.~Tarantino,
  arXiv:0807.4605 [hep-lat].

\bibitem{Abulencia:2006ze}
  A.~Abulencia {\it et al.}  [CDF Collaboration],
  Phys.\ Rev.\ Lett.\  {\bf 97}, 242003 (2006)
  [arXiv:hep-ex/0609040].

\bibitem{D0Coll:Bsmixing}
  D0 Collaboration, D0 Note 5618 -- CONFv1.2.

\bibitem{:2008cy}
  B.~Aubert {\it et al.}  [BABAR Collaboration],
  arXiv:0808.1915 [hep-ex].

\bibitem{:2008gf}
  B.~Aubert {\it et al.}  [BABAR Collaboration],
  arXiv:0808.1379 [hep-ex].

\bibitem{Taniguchi:2008ty}
  N.~Taniguchi, M.~Nakao, S.~Nishida and f.~t.~B.~Collaboration,
  arXiv:0804.4770 [hep-ex].

\bibitem{:2008gvb}
  B.~Aubert  {\it et al.}  [BABAR Collaboration],
  arXiv:0805.4796 [hep-ex].

\bibitem{Anisimovsky:2004hr}
  V.~V.~Anisimovsky {\it et al.}  [E949 Collaboration],
  Phys.\ Rev.\ Lett.\  {\bf 93}, 031801 (2004)
  [arXiv:hep-ex/0403036].

\bibitem{Biino:2008ja}
  C.~Biino and M.~Pepe,
  arXiv:0809.4969 [hep-ex].

\bibitem{Acosta:2005hr}
  D.~E.~Acosta {\it et al.}  [CDF Collaboration],
  Phys.\ Rev.\ Lett.\  {\bf 95}, 102002 (2005)
  [arXiv:hep-ex/0505091].

\bibitem{Abazov:2008yn}
  V.~M.~Abazov {\it et al.}  [D0 Collaboration],
  Phys.\ Rev.\ Lett.\  {\bf 100}, 192003 (2008)
  [arXiv:0801.1326 [hep-ex]].


\bibitem{Abazov:2006gd}
  V.~M.~Abazov {\it et al.}  [D0 Collaboration],
  Phys.\ Rev.\ Lett.\  {\bf 98}, 181802 (2007)
  [arXiv:hep-ex/0612052].

\bibitem{Aaltonen:2008sy}
  T.~Aaltonen {\it et al.}  [CDF Collaboration],
  arXiv:0809.2581 [hep-ex].


\end{thebibliography}
\end{document}